\documentclass[aps,prb,amsmath,amssymb,reprint,superscriptaddress,preprintnumbers,showpacs,intlimits,longbibliography]{revtex4-2}
\pdfoutput=1

\usepackage{bm,latexsym,mathrsfs,enumerate,revsymb}
\usepackage{mathtools,upgreek}
\usepackage[colorlinks = true,breaklinks=true,unicode=true,urlcolor = blue,citecolor = blue,linkcolor = blue]{hyperref}
\usepackage{graphicx}
\usepackage{multirow}
\renewcommand{\vec}[1]{\bm{#1}}
\newcommand{\Hquad}{\hspace{0.5em}}
\newcommand{\ts}[1]{\textsc{#1}}
\newcommand{\dd}{\mathrm{d}}

\newcommand{\x}{\mathcal{X}} 

\usepackage[final]{pdfpages}
\usepackage{tikz}
\makeatletter
\AtBeginDocument{\let\LS@rot\@undefined}
\makeatother

\begin{document}
%======================================================================================================================
%============================================TITLE=====================================================================
%======================================================================================================================
\title{Dynamics of domain walls in curved antiferromagnetic wires}
%======================================================================================================================
%============================================AUTHOR====================================================================
%======================================================================================================================
\author{Kostiantyn V. Yershov}
\email{yershov@bitp.kiev.ua}
\affiliation{Leibniz-Institut f{\"u}r Festk{\"o}rper- und Werkstoffforschung, IFW Dresden, 01069 Dresden, Germany}
\affiliation{Bogolyubov Institute for Theoretical Physics of the National Academy of Sciences of Ukraine, 03143 Kyiv, Ukraine}
%======================================================================================================================
%======================================================================================================================
%======================================================================================================================	
\date{\today}
%======================================================================================================================
%============================================ABSTRACT==================================================================
%======================================================================================================================
\begin{abstract}
	The influence of the curvature on the dynamical properties of transversal domain walls in a thin antiferromagnetic wire is studied  theoretically. Equations of motion for antiferromagnetic domain wall are obtained within the collective variable approach ($q-\Phi$ model). It is shown that (i) for the case of localized bend, curvature results in a pinning potential for domain wall. (ii)~The gradient of the curvature results in a driving force on the domain wall and it effectively moves without any external stimuli. Although we showcase our approach on the specific parabola and Euler spiral geometries, the approach is general and valid for a wide class of geometries. All analytical predictions are confirmed by numerical simulations.
\end{abstract}

\maketitle

\textit{Introduction. -- } Antiferromagnets are promising candidates for novel spintronic devices~\cite{Jungwirth16,Baltz18,Gomonay18,Wadley16,Kosub17}. One of the key elements in these devices are domain walls (DWs). DWs are considered as carriers of essential information on the magnetic microstructure of a material~\cite{Hubert09,Weber03} and logical bits in magnetic memory devices~\cite{Parkin08}. Furthermore, and similarly to ferromagnet-based DW logic~\cite{Allwood05}, the understanding and control of antiferromagnetic (AFM) DWs could inform future approaches to AFM spintronics architectures.

Recent progress in the field of AFM spintronics has been made on the deterministic manipulation of AFM DWs using different external stimuli. AFM DWs can be effectively moved by using spin-transfer torque~\cite{Swaving11,Hals11,Tveten13,Yamane17}, spin-orbit torque~\cite{Gomonay16a,Shiino16,Sanchez-Tejerina20}, rotating magnetic field~\cite{Pan18}, and by propagation of spin waves~\cite{Kim14a,Tveten14}. All these studies are considering the DW dynamics in rectilinear systems, while role of geometrical curvature remains unclear and general theory for AFM DW dynamics in curved systems is absent. Nevertheless, one should note some progress in theoretical studying of curvature-induced effects on equilibrium states in rings~\cite{Castillo-Sepulveda17,Pylypovskyi21f} and helices~\cite{Pylypovskyi20}.

Here, we demonstrate that a gradient of curvature of the AFM wire can be considered as a driving force for AFM DW, i.e. AFM DW moves without any external stimuli. We show that (i) a localized bend of a wire results in a pinning potential for a DW; (ii) DW performs a translational motion under the action of the gradient of the curvature. The static properties of AFM DWs are similar to those of ferromagnetic (FM) DW~\cite{Yershov15b,Yershov16,Pylypovskyi16}, while the dynamics is fundamentally different. The proposed approach is general and valid for a wide class of geometries. We applied our approach for the specific parabola and Euler spiral geometries, which cover any local geometry that one wants to cover. The analytical results are confirmed with numerical simulations.

\textit{Model. -- }We consider a thin wire made from an intrinsically achiral two-sublattice AFM material. The magnetic moments $\vec{\mu}_i=\vec{M}_i/M_0$ are arranged along the space curve~$\vec{\gamma}(s)$ with $s$ being an arc length coordinate, $M_0$ is the length of magnetic moment, and index $i$ enumerates sublattices. The curved 1D wire~$\vec{\gamma}(s)$ lies within the $xy$ plane and can be parameterized with curvature $\kappa(s)$. The local reference frame can be chosen as the Frenet--Serret frame with tangential $\vec{e}_\textsc{t} = \partial_s \vec{\gamma}$, normal $\vec{e}_\textsc{n} = \partial_s \vec{e}_\textsc{t} / \kappa$, and binormal vectors $\vec{e}_\textsc{b} = \vec{e}_\textsc{t} \times \vec{e}_\textsc{n}$. The order parameters of antiferromagnet are given by the N{\'e}el vector $\vec{n} = \left(\vec{\mu}_1 - \vec{\mu}_2\right)/2$ and total magnetization $\vec{m} = \left(\vec{\mu}_1 + \vec{\mu}_2\right) / 2$, where $\vec{m}\cdot\vec{n} = 0$ and $\vec{n}^2 + \vec{m}^2 = 1$. The order parameters are assumed to be functions of a single spatial coordinate $s$. The model, thus, describes wires in the limit of very thin but fully compensated AFM structures, therefore now inhomogeneity is considered in the normal $\vec{e}_\textsc{n}$ and binormal $\vec{e}_\textsc{b}$ directions.

In the following, we will consider the case of a strong exchange field $H_\textsc{x}$ acting between the magnetic sublattices~\footnote[1]{Action of the exchange field $H_\textsc{x}$ can be expressed in terms of the uniform exchange interaction  between two sublattices. The interaction energy density is $\mathscr{E}_\textsc{x}^\textsc{u} =H_\ts{x}M_0\left(\vec{\mu}_1\cdot\vec{\mu}_2\right)$.}. It is large compared to the anisotropy field $H_\textsc{a}$, i.e.~$\sqrt{H_\textsc{a}/H_\textsc{x}} = \zeta \ll 1$. In this case, the magnetization is small ($|\vec{m}|\ll 1$ and $|\vec{n}| \approx 1$), and the state of the AFM is determined by the N{\'e}el vector. The magnetic properties of the system can be modeled by means of the energy density
\begin{equation}\label{eq:model}
	\mathscr{E} = A \left(\partial_{s}\vec{n}\right)^2 + H_\textsc{a} M_0 \left[1 - \left(\vec{n}\cdot\vec{e}_\textsc{t}\right)^2\right],
\end{equation}
which is valid for antiferromagnets if the external magnetic field is absent. The first term in~\eqref{eq:model} describes the exchange interaction with the exchange stiffness constant $A$, the second term is the easy-tangential anisotropy with easy-axis $\vec{e}_\textsc{t}$ being tangential to the wire.

The dynamics of the N{\'e}el vector can be described within the Lagrange formalism~\cite{Baryakhtar80,Kosevich90,Ivanov05a,Qaiumzadeh18}. The Lagrangian $\mathcal{L}$ and Rayleigh dissipation function $\mathcal{R}$ normalized by $\sqrt{AH_\textsc{a}M_0}$ are given by
\begin{subequations}\label{eq:lagrangian}
	\begin{equation}\label{eq:lagrangian_n}
		\mathcal{L} = \int_{-\infty}^{+\infty}\dot{\vec{n}}^2\dd\xi - \mathcal{E},\quad \mathcal{R} = \frac{\eta}{\zeta}\int_{-\infty}^{+\infty}\dot{\vec{n}}^2\dd\xi
	\end{equation}
	with total energy written in curvilinear frame of reference~\cite{Pylypovskyi20}
	\begin{equation}\label{eq:energy_tnb}
		\mathcal{E} = \!\!\int_{-\infty}^{+\infty}\!\!\bigr[n_\alpha'n_\alpha' + \mathcal{D}_{\alpha\beta}\!\left(n_\alpha'n_\beta - n_\alpha n_\beta'\right)\! + \mathcal{K}_{\alpha\beta}n_\alpha n_\beta\bigl]\dd\xi.
\end{equation}
\end{subequations}
Here, overdot indicates derivatives with respect to the dimensionless time $\uptau = \omega_0 t$ with $\omega_0 = \gamma_0\sqrt{H_\textsc{a}H_\textsc{x}}$ being the frequency of uniform AFM resonance and $\gamma_0$ being gyromagnetic ratio, prime denotes the derivative with respect to the dimensionless arc length $\xi = s /\ell$ with $\ell = \sqrt{A / \left(H_\textsc{a}M_0\right)}$ being the length scale of the system which defines the width of DW in rectilinear system, $\eta$ is a damping coefficient, and $n_\alpha$ is a curvilinear component of the N{\'e}el vector. Here, Greek indexes run over the components of the $\left\{\textsc{t},\textsc{n},\textsc{b}\right\}$ frame. The first term in energy~\eqref{eq:energy_tnb} corresponds to the common inhomogeneous exchange interaction in rectilinear system, the second term is a curvature-induced Dzyaloshinskii--Moriya interaction with $\mathcal{D}_{\alpha\beta}$ being the effective DMI tensor with two non-zero coefficients $\mathcal{D}_{\textsc{t}\textsc{n}} = -\mathcal{D}_{\textsc{n}\textsc{t}} = \varkappa$ and $\varkappa = \kappa\ell$ is the dimensionless curvature. The last term is an effective anisotropy $\mathcal{K}_{\alpha\beta} = \mathcal{D}_{\alpha\gamma}\mathcal{D}_{\beta\gamma} - \delta_{\textsc{t}\alpha}\delta_{\textsc{t}\beta}$ with $\delta_{\alpha\beta}$ being the Kronecker delta. The specific form of $\mathcal{D}_{\alpha\beta}$ takes into account that a plane wire has zero torsion.

A small magnetization arises due to dynamics of the N{\'e}el vector:
\begin{equation}\label{eq:magnetization}
	\vec{m} = \zeta\left[\dot{\vec{n}}\times \vec{n}\right].
\end{equation}
Derivation of Eq.~\eqref{eq:magnetization} and procedure of exclusion of the magnetization from the AFM dynamics can be found in a number of previous works, for details see Refs.~\cite{Baryakhtar80,Ivanov95f,Gomonay10}.

In the following we will utilize the assumption that the length of $\vec{n}$ is constant and we will use the angular parameterization of N{\'e}el order parameter $\vec{n} = \vec{e}_\textsc{t}\cos\theta + \sin\theta \left(\vec{e}_\textsc{n}\cos\phi + \vec{e}_\textsc{b}\sin\phi\right)$.

\textit{Collective variable approach. -- }In order to study the statics and dynamics of the DW in curved AFM system we will use a collective variable approach~\cite{Slonczewski72,Thiele73} based on the simple $q-\Phi$ model~\cite{Slonczewski72,Malozemoff79}
\begin{equation}\label{eq:q-Phi}
	\cos\theta = -p \tanh \frac{\xi-q(\uptau)}{\Delta}, \quad \phi = \Phi(\uptau).
\end{equation}
Here, $q$ and $\Phi$ are time-dependent collective variables, which determine the DW position and phase~(orientation of the transversal componenet of $\vec{n}$), respectively; $p$ is a topological charge, which determines the DW type: head-to-head ($p =+1$) or tail-to-tail ($p = -1$). The DW width $\Delta$ is assumed to be a slaved variable~\cite{Hillebrands06,Landeros10,Yershov18a}, i.e., $\Delta(\uptau) = \Delta\left[q(\uptau),\Phi(\uptau)\right]$. The model~\eqref{eq:q-Phi} coincides with the exact DW solution for wires with constant curvature~($\varkappa = \text{const}$ -- circle, $\varkappa = 0$ -- rectilinear wire). In the following, the curvature is considered as a small perturbation, which results in a driving force while keeping the form~\eqref{eq:q-Phi} unchanged.

%======================================================================================================================
%														FIGURE 1
%======================================================================================================================
\begin{figure}[t]
	\includegraphics[width=\columnwidth]{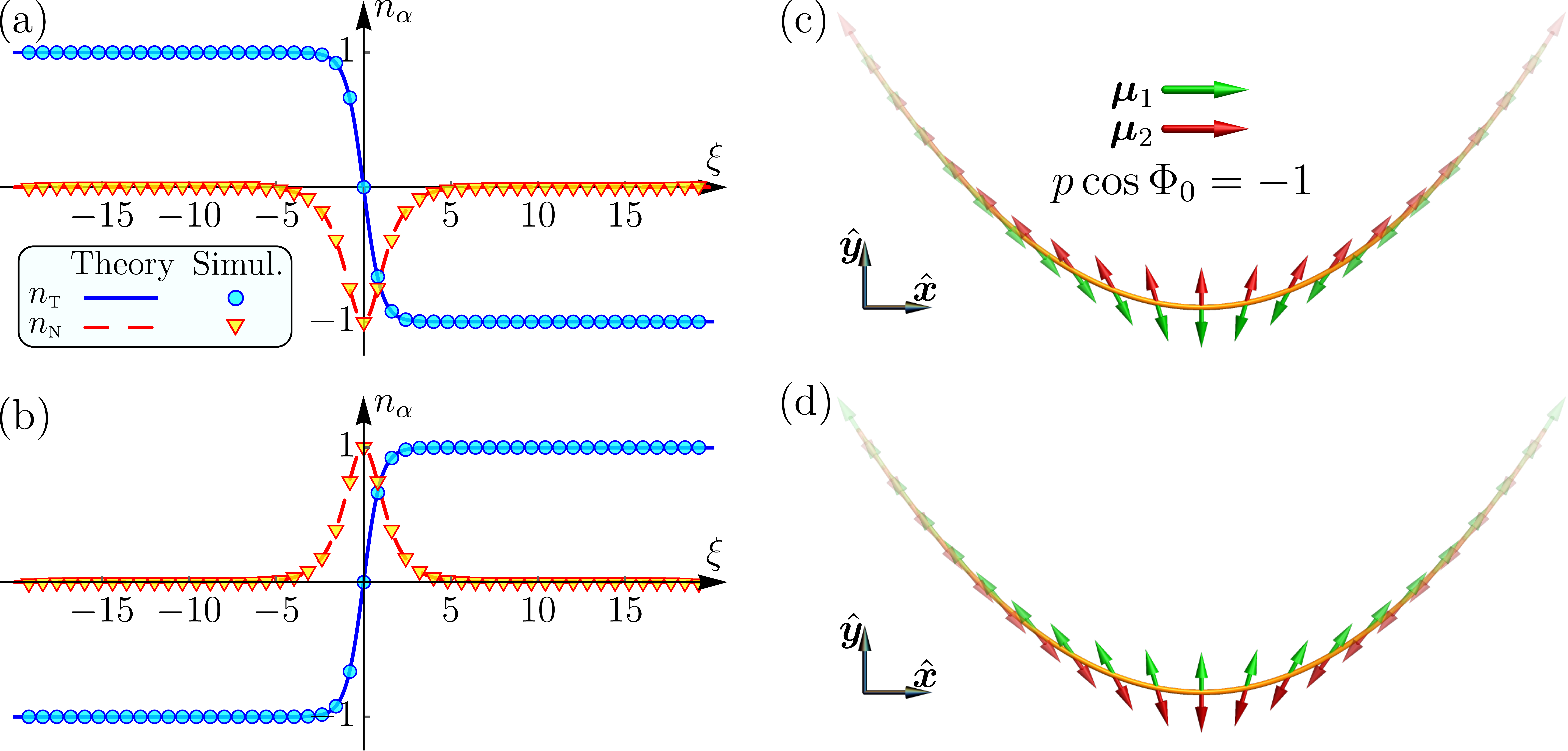}
	\caption{(Color online) Transversal AFM DW at the parabolic wire bend with $\varkappa_0 = 0.05$ for: (a) $p=+1$ and $\cos\Phi_0 = - 1$; (b) $p = -1$  and $\cos\Phi_0 = + 1$. Symbols and lines in (a) and (b) correspond to the data obtained by numerical simulations~\cite{Note2} and Ansatz~\eqref{eq:q-Phi}, respectively. (c) and (d) show the distribution of magnetic moments of two sublattices for the stable DW structure in the parabolic wire.}\label{fig:dw}
\end{figure}
%======================================================================================================================
%======================================================================================================================

%======================================================================================================================
%														FIGURE 2
%======================================================================================================================
\begin{figure*}[t]
	\includegraphics[width=\textwidth]{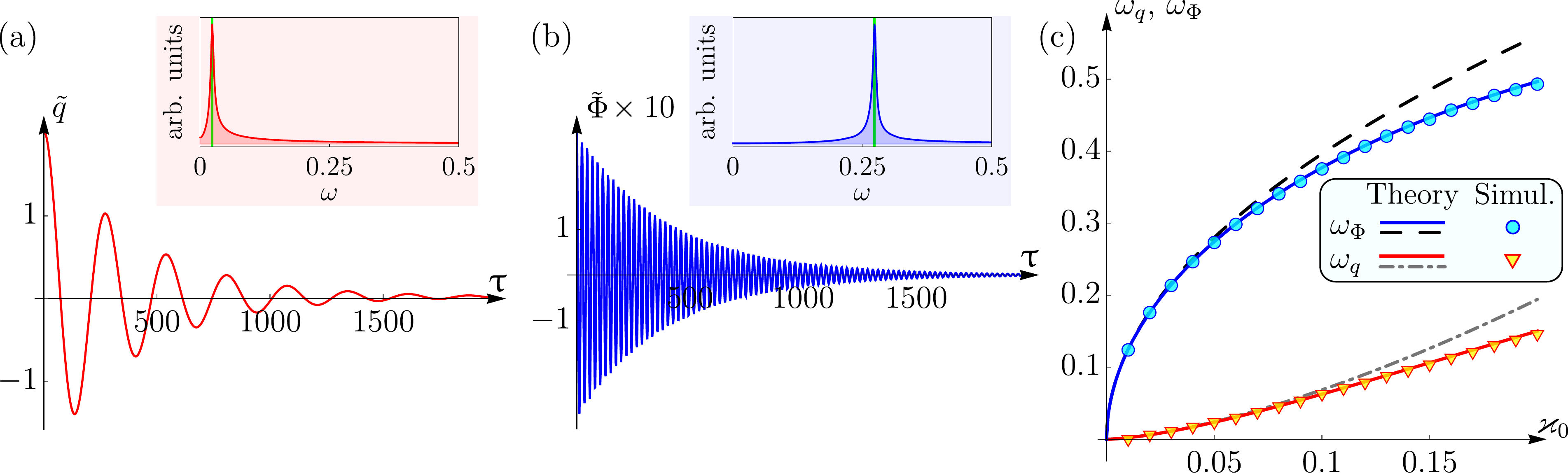}
	\caption{(Color online) Dynamics of the transversal DW in a parabolic wire bend. (a) and (b) show the time dependencies~(with the corresponding Fourier spectra) of position $\tilde{q} = q-q_0$ and phase $\tilde{\Phi} = \Phi - \Phi_0$, respectively, for wire with $\varkappa_0 = 0.05$. (c) Eigenfrequencies of the DW position and DW phase oscillations in vicinity of the equilibrium. Dashed and solid lines in (c) correspond to the predictions~\eqref{eq:qPhi-omega1} and \eqref{eq:qPhi-omega2}, respectively. Symbols show the results of numerical simulations. In all simulations we have $\eta = 10^{-4}$ and $\zeta =2\times 10^{-2}$, for details see supplemental material~\cite{Note2}.}\label{fig:q_Phi_osc}
\end{figure*}
%======================================================================================================================
%======================================================================================================================

\textit{Statics of AFM DW in curved wire. -- } Substituting the Ansatz~\eqref{eq:q-Phi} into~\eqref{eq:energy_tnb} and performing integration over the arc length $\xi$, we obtain the energy in the form
\begin{equation}\label{eq:E-q-Phi}
	\mathcal{E}^\textsc{dw} = \mathcal{E}_0 + 2p\pi\varkappa(q) \cos\Phi - 2\varkappa^2(q)\Delta\sin^2\Phi.
\end{equation}
The first term $\mathcal{E}_0 = 2\left(1+\Delta^2\right)/\Delta$ in~\eqref{eq:E-q-Phi} corresponds to the energy of the DW in a rectilinear system, and determines competition of the inhomogeneous exchange and anisotropy contributions. Terms linear and quadratic with respect to curvature correspond to the geometry-induced DMI and anisotropy driven by the exchange~\cite{Sheka15,Pylypovskyi20}, respectively. Term linear with respect to $\varkappa$ in the  energy~\eqref{eq:E-q-Phi} demonstrates the coupling between the curvature, DW topological charge, and phase, i.e. DW energy is minimized when $\cos\Phi_0 = -\text{sgn}(p\varkappa)$ and such a DW wall is referred as a stable. One should note that the N{\'e}el vector is a director, and DWs with $\left\{p=+1,\cos\Phi_0 = -1\right\}$ and $\left\{p=-1,\cos\Phi_0 = +1\right\}$ for $\varkappa>0$ are equivalent~\footnote[3]{One should obtain a similar result for the case of negative curvature $\varkappa<0$. In this case the stable DW has $p \cos\Phi_0 = +1$, i.e. stable DWs are realized for $\left\{p=+1,\cos\Phi_0 = +1\right\}$ and $\left\{p=-1,\cos\Phi_0 = -1\right\}$.}, for details see Fig.~\ref{fig:dw}. From~\eqref{eq:E-q-Phi} we conclude that the position of stable equilibrium $q_0$ for DW is determined by condition
\begin{equation}\label{eq:equilibrium_q}
	\varkappa'(q_0) = 0.
\end{equation}
Condition~\eqref{eq:equilibrium_q} means that the stable DW with $p\cos\Phi_0=-1$ ($p\cos\Phi_0=+1$) has minimal energy at the maximal (minimal) value of curvature. We expect that deviations of the DW position and phase from equilibrium values will result in the curvature-induced dynamics for DW. The effect of curvature-induced motion will be discussed further. 

The minimization of the energy~\eqref{eq:E-q-Phi} with respect to the DW width results in the equilibrium value $\Delta_0 = 1$, which is the same as for rectilinear wire. However, if values of the DW position and phase deviate from equilibrium, the DW width is defined as $\Delta(\uptau)= 1/\sqrt{1-\varkappa^2(q)\sin^2\Phi}$. Here, coefficient $\varkappa^2(q)$ acts as an easy-binormal anisotropy constant.

\textit{Curvature-induced dynamics of the AFM DW. -- }Let us now proceed to dynamical properties of the AFM DW. The Lagrangian and Rayleigh dissipation function~\eqref{eq:lagrangian_n} in terms of the collective variables can be written as
\begin{equation}\label{eq:LR-q-Phi}
	\begin{split}
		\mathcal{L}^\textsc{dw} =&\frac{2}{\Delta}\left[\dot{q}^2+\left(\dot{\Phi}\Delta\right)^2\right]-\mathcal{E}^\textsc{dw},\\
		\mathcal{R}^\textsc{dw} =&\frac{2}{\Delta}\frac{\eta}{\zeta}\left[\dot{q}^2+\left(\dot{\Phi}\Delta\right)^2\right].
	\end{split}
\end{equation}
It is well known, that for FM DWs the collective variables $q$ and $\Phi$ are conjugated variables~\cite{Hillebrands06}, i.e. dynamics of $q$ induces dynamics of $\Phi$ and vice versa. Additionally, for FM DW, the phase is canonically conjugated momentum to the DW position. This is not the case for AFM DW, where DW position and phase have their own momentum.

By substituting~\eqref{eq:LR-q-Phi} into the Lagrange-Rayleigh equations one obtains
\begin{equation}\label{eq:q-Phi-motion}
	\begin{split}
		\frac{4}{\Delta}\left[\ddot{q}+\frac{\eta}{\zeta}\dot{q}\right] = -\frac{\partial\mathcal{E}^\textsc{dw}}{\partial q},\\
		4\Delta\left[\ddot{\Phi}+\frac{\eta}{\zeta}\dot{\Phi}\right] = -\frac{\partial\mathcal{E}^\textsc{dw}}{\partial \Phi}.
	\end{split}
\end{equation}
The second equation of the set~\eqref{eq:q-Phi-motion} has a general solution~$\cos\Phi_0 = \pm 1$. It means that during dynamics the transversal part of DW remains within the wire plane. The constant DW phase has intuitive explanation: during the curvature-induced drift DW moves to the area with bigger curvature. In this case, the term linear with respect to curvature in the energy~\eqref{eq:E-q-Phi} becomes dominant, which results in the fixing of phase $\Phi_0=0$ or $\Phi_0=\pi$.

Firstly, we will consider dynamics of the AFM DW in the wire with localized curvature $\varkappa\left(\pm\infty\right) = 0$. Here we are interested in linear dynamics of the DW in the vicinity of the equilibrium state. Therefore, we introduce small deviations in the way $q(\uptau) = q_0 + \tilde{q}(\uptau)$ and $\Phi(\uptau) = \Phi_0 + \tilde{\Phi}(\uptau)$. For the limit of weak curvature ($\varkappa\ll 1$) the equations of motion~\eqref{eq:q-Phi-motion} linearized with respect to the deviations  read
\begin{equation}\label{eq:q-Phi-motion-linear}
	\begin{split}
		\ddot{\tilde{q}}&+\frac{\eta}{\zeta}\dot{\tilde{q}} \approx \frac{\pi}{2}\varkappa''(q_0)\tilde{q},\\
		\ddot{\tilde{\Phi}}&+\frac{\eta}{\zeta}\dot{\tilde{\Phi}} \approx -\frac{\pi}{2}\varkappa(q_0)\tilde{\Phi}.
	\end{split}
\end{equation}
The solution of~\eqref{eq:q-Phi-motion-linear} results in the decaying oscillations $\tilde{q}(\uptau) = A_{\tilde{q}}\cos\left(\omega_q \uptau + \psi_{\tilde{q}}\right)e^{-\eta \uptau/\left(2\zeta\right)}$ and $\tilde{\Phi}(\uptau) = A_{\tilde{\Phi}}\cos\left(\omega_\Phi \uptau + \psi_{\tilde{\Phi}}\right)e^{-\eta \uptau/\left(2\zeta\right)}$ with frequencies~(for the case of low damping $\eta\ll1$)
\begin{equation}\label{eq:qPhi-omega1}
	\omega_q \approx \sqrt{\pi\left|\varkappa''(q_0)\right|/2},\quad \omega_\Phi \approx \sqrt{\pi\left|\varkappa(q_0)\right|/2}.
\end{equation}
The phases $\psi_{\tilde{q}}$ and $\psi_{\tilde{\Phi}}$ are determined by the initial conditions. Remarkably, for the case of a circular wire segment ($\varkappa = \text{const}$), where curvature does not produce any geometrical pinning potential, the DW phase has nonzero frequency $\omega_{\Phi}$ while the position evolves as $\tilde{q} - \tilde{q}_0 = v_0\zeta\left(1-e^{-\eta\,\uptau/\zeta}\right)/\eta$, where $\tilde{q}_0$ and $v_0$ are an initial DW position and velocity, respectively.

In the more general case~$\varkappa \in \left[0;1\right]$ the Ansatz~\eqref{eq:q-Phi} leads to the following expressions for the eigenfrequencies, for details see Supplemental Materials~\cite{Note2},
\begin{equation}\label{eq:qPhi-omega2}
	\begin{split}
		\omega_q &\approx \sqrt{\frac{|\mathcal{I}_q|}{2}},\Hquad \omega_\Phi \approx \sqrt{\frac{|\mathcal{I}_\Phi|}{2}},\Hquad \mathcal{I}_q = \!\!\int_{-\infty}^{+\infty}\!\!\frac{\varkappa''\left(\xi\right)}{\cosh\left(\xi-q_0\right)}\mathrm{d}\xi,\\
		\mathcal{I}_\Phi&= \int_{-\infty}^{+\infty}\left[\frac{\varkappa(\xi)}{\cosh\left(\xi-q_0\right)}-\frac{\varkappa^2(\xi)}{\cosh^2\left(\xi-q_0\right)}\right]\mathrm{d}\xi.
	\end{split}
\end{equation}
In this case the equilibrium position is defined by the equation $\int_{-\infty}^{+\infty}\mathrm{d}\xi\,\varkappa'(\xi)/\cosh\left(\xi-q_0\right)=0$. The corresponding frequencies are presented in Fig.~\ref{fig:q_Phi_osc}.

In order to check general results obtained above, we performed a set of numerical simulations~\cite{Note2} for the case of parabolic wires with geometry
\begin{equation}\label{eq:geo_parabola}
	\vec{\gamma}(\xi) = x(\xi)\hat{\vec{x}} + \varkappa_0 \frac{x^2(\xi)}{2}\hat{\vec{y}}.
\end{equation}
For the parabolic wire~\eqref{eq:geo_parabola} the equilibrium position is $q_0=0$, which corresponds to the extreme value of the curvature $\varkappa_0$. In the limit of weak curvature the resulting frequencies of oscillations for DW position and phase are $\omega_q = \sqrt{3\pi |\varkappa_0^3|/2}$ and $\omega_\Phi = \sqrt{\pi|\varkappa_0|/2}$, respectively, see~Fig.~\ref{fig:q_Phi_osc}. One can see, that frequency of the DW position oscillations is of third order of magnitude with respect to the curvature compared to the DW phase oscillations.

%======================================================================================================================
%														FIGURE 3
%======================================================================================================================
\begin{figure}[t]
	\includegraphics[width=0.8\columnwidth]{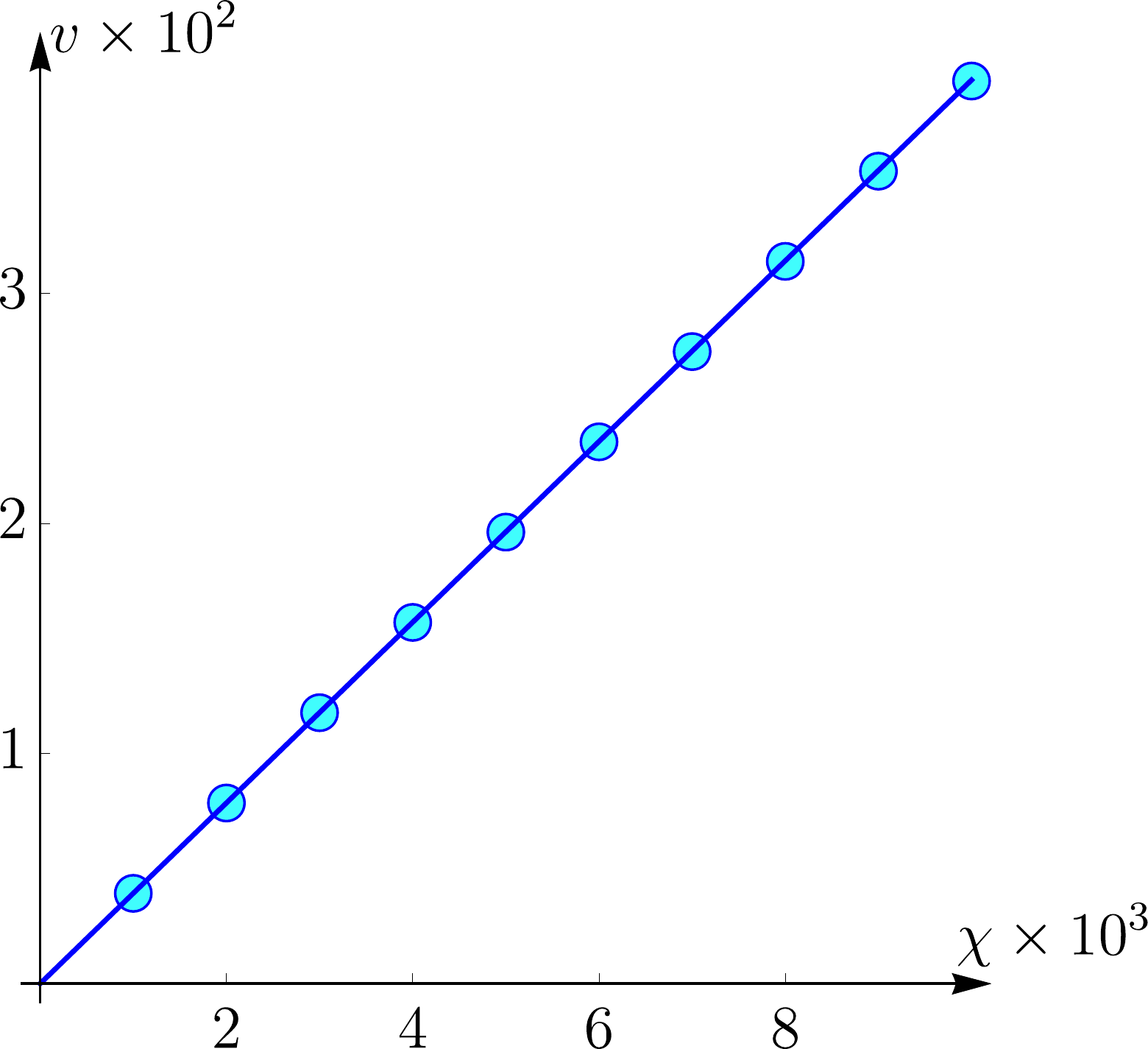}
	\caption{(Color online) Asymptotic DW velocity $v$ as a function of the gradient of the curvature $\chi$. Solid line gives the analytical prediction~\eqref{eq:dw_velocity}. Symbols show the results of numerical simulations. In all simulations we have $\eta = 10^{-2}$ and $\zeta =2.5\times 10^{-2}$, for details see supplemental materials~\cite{Note2}.}\label{fig:v_vs_kappa}
\end{figure}
%======================================================================================================================
%======================================================================================================================

In a second step, the general equation of motion~\eqref{eq:q-Phi-motion} are applied for the case of the Euler spiral geometry~\cite{Lawrence14}, also known as Cornu spiral or clothoid. The geometry of the Euler spiral is defined as
\begin{equation}\label{eq:geo_Euler}
	\vec{\gamma}(\xi) =\sqrt{\pi/\chi} \left[\mathrm{C}(\xi\sqrt{\chi/\pi})\hat{\vec{x}} + \mathrm{S}(\xi\sqrt{\chi/\pi})\hat{\vec{y}}\right],
\end{equation}
where $\mathrm{C}(u) =\int_{0}^u\cos\left[\pi x^2/2\right]\mathrm{d}x$ and $\mathrm{S}(u) =\int_{0}^u\sin\left[\pi x^2/2\right]\mathrm{d}x$ are Fresnel integrals. The curvature of this curve is a linear function of the arc length $\varkappa = \chi\xi$ with $\chi$ being a dimensionless gradient of the curvature. 

For the Euler spiral, equations of motion~\eqref{eq:q-Phi-motion} results in a constant DW phase $\Phi(\uptau) = \Phi_0$, while DW position moves with velocity
\begin{equation}\label{eq:dw_velocity}
	\begin{split}
		\dot{q}(\uptau) = v + \left[v_0-v\right]e^{-\eta \uptau/\zeta},\quad\! v=\chi\frac{\pi}{2}\frac{\zeta}{\eta},
	\end{split}
\end{equation}
where $v_0$ is an initial DW velocity and $v = \dot{q}\left(\uptau\to\infty\right)$ is an asymptotic velocity for DW. In a short time limit $\uptau\ll\zeta/\eta$ or vanishing damping~($\eta\to 0$) we can approximate DW velocity as $\dot{q} \approx v_0 + a \uptau$ with $a=\pi\chi/2$ being a DW acceleration. Here we can see that DW velocity increases linearly with time.  The resulting DW velocity~\eqref{eq:dw_velocity} as a function of the gradient of the curvature is plotted in Fig.~\ref{fig:v_vs_kappa}.  One should note that for velocities close to magnons velocity the model should be revised by considering deformations of the DW shape.

It is instructive to compare the curvature-induced effects for FM and AFM DWs in curved wires. The curvature-induced driving for DWs in FM and AFM wires originate from exchange-driven Dzyaloshinskii--Moriya interaction~\cite{Sheka15,Pylypovskyi20}. For both cases of magnetic ordering, in wires with localized curvature DWs are pinned at the maximal value of curvature distribution, which also results in phase selectivity for DWs~\cite{Kim14,Yershov15b,Yershov16}. While the statics for different ordering remain the same behavior, the dynamics is essentially different. (i) For FM DWs the eigenfrequency oscillations of position and phase, in the vicinity of equilibrium, are the same $\omega_{q,\Phi}^\textsc{fm}\propto\sqrt{|\varkappa(q_0)\varkappa''(q_0)|}$~\cite{Yershov15b}, while for AFM DWs they are different, see Eq.~\eqref{eq:qPhi-omega1}. Remarkably, for the case of constant curvature~(the case of circular wire segment) one can observe oscillations of the phase for AFM DW with finite frequency $\omega_{\Phi}\propto\sqrt{|\varkappa(q_0)|}$. (ii) For the AFM DW we obtained constant asymptotic velocity $v\propto \chi$ and phase $\Phi = \text{const}$, while in uniaxial FM wire DWs move with velocity which increases exponentially. The constant asymptotic velocity for FM DWs is possible only in biaxial wires where phase behaves as $\Phi^\textsc{fm}-\Phi_0\propto 1/q^\textsc{fm}(\uptau)$~\cite{Yershov18a}. (iii) In a limit case of zero damping ($\eta =0$) AFM DWs move with constant acceleration $a \propto \chi$ and $\Phi=\text{const}$, while FM DWs move with velocity $v^\textsc{fm}\propto \chi e^{\chi \uptau}$ and $\Phi^\textsc{fm}_{\uptau\to\infty}\to \Phi_0 -\pi/2  \cos\Phi_0$.

\textit{Conclusions. -- }In conclusion, we demonstrate the effect of curvature-induced pinning and driving for the transversal DW in thin planar AFM wires. The origin of the pinning/driving is exchange-driven effective DMI~\cite{Sheka15,Pylypovskyi20}. We obtain expressions for eigenfrequencies~\eqref{eq:qPhi-omega2} of DW oscillations in vicinity of the equilibrium state. For the case of weak curvature the approximation~\eqref{eq:qPhi-omega1} can be used, see Fig.~\ref{fig:q_Phi_osc}(c). Curvature-induced motion of the DW is accompanied by constant DW phase~($\cos\Phi_0 = \pm 1$), while DW velocity is defined by the gradient of the curvature $v\propto \chi$, see Eq.~\eqref{eq:dw_velocity}.

In a linear approximation with respect to the curvature, presented approach can be applied for the arbitrary curved non-planar wires with small torsion. In this case torsion results in negligibly small corrections of the second order of magnitude.

\textit{Acknowledgments. -- }The author is grateful to Dr.~V.~Kravchuk and Dr.~U.~R{\"o}{\ss}ler for fruitful discussions, and U. Nitzsche for technical support. In part this work was supported by the National Research Foundation of Ukraine (Project No. 2020.02/0051).



\section*{Supplementary material}

# Supplementary to
# "Dynamics of domain walls in curved antiferromagnetic wires"

Kostiantyn V. Yershov[1, 2, *]

[1]*Leibniz-Institut für Festkörperforschung, IFW Dresden, 01069 Dresden, Germany*
[2]*Bogolyubov Institute for Theoretical Physics of the National Academy of Sciences of Ukraine, 03143 Kyiv, Ukraine*

The supplementary information provides details on analytical calculations of main aspect of the transversal domain wall dynamics in a wire with non-zero gradient of the curvature.

## I. DETAILS OF THE $q - \Phi$ MODEL FOR $\varkappa < 1$

Substituting the Ansatz (4) into the energy expression (2b) and performing the integration one obtains (up to an additive constant)

$$\frac{\mathcal{E}^{\text{DW}}}{2} = \frac{\mathcal{E}_0}{2} + \frac{1}{\Delta^2}\left[p\,\mathcal{W}_1 \cos\Phi - \mathcal{W}_2 \sin^2\Phi\right],$$
$$\mathcal{W}_n = \int\limits_{-\infty}^{+\infty} \frac{w^n(\xi)}{n}\mathrm{d}\xi, \quad w(\xi) = \varkappa\Delta/\cosh\left[\frac{\xi-q}{\Delta}\right]. \tag{S1}$$

In the limit case $\varkappa \ll 1$ the expression (S1) is reduced to (5). For the more general case $0 < \varkappa < 1$ one has $w_1 < 1$ and consequently $\mathcal{W}_1 > \mathcal{W}_2$. In this case the value of phase $\Phi$ which minimizes the energy (S1) is determined as $\cos\Phi_0 = -p$ and the corresponding value of the equilibrium domain wall position $q_0$ is determined as

$$\int\limits_{-\infty}^{+\infty} \frac{\varkappa'(\xi)}{\cosh[\xi - q_0]}\mathrm{d}\xi = 0. \tag{S2}$$

The equations of motion linearized in the vicinity of the equilibrium $q_0$ and $\Phi_0$ read

$$\ddot{\tilde{q}} + \frac{\eta}{\zeta}\dot{\tilde{q}} \approx -\tilde{q}\frac{\mathcal{I}_q}{2},$$
$$\ddot{\tilde{\Phi}} + \frac{\eta}{\zeta}\dot{\tilde{\Phi}} \approx -\tilde{\Phi}\frac{\mathcal{I}_\Phi}{2}. \tag{S3}$$

Constants $\mathcal{I}_q$ and $\mathcal{I}_\Phi$ are defined in (11).

## II. DETAILS ON NUMERICAL SIMULATIONS

In order to verify our analytical calculations we perform a set numerical simulations for AFM curved wires. We consider a single chain with lattice constant $a$. Each node is characterized by a magnetic moment $\boldsymbol{m}_i(t)$ which is located at the position $\boldsymbol{r}_i$. Here $i \in \mathbb{N}$ defines the magnetic moment and its position on the chain with size

* yershov@bitp.kiev.ua

$i \in [1, N]$. The dynamics of magnetic system is govern by discrete Landau–Lifshitz–Gilbert equations

$$\frac{\mathrm{d}\boldsymbol{m}_i}{\mathrm{d}t} = \frac{\gamma_0}{\mu_s}\left[\boldsymbol{m}_i \times \frac{\partial \mathscr{H}}{\partial \boldsymbol{m}_i}\right] + \eta\left[\boldsymbol{m}_i \times \frac{\mathrm{d}\boldsymbol{m}_i}{\mathrm{d}t}\right], \tag{S4}$$

where $\mu_s$ is a magnetic moment of a magnetic site. The Hamiltonian of a magnetic system has following form

$$\mathscr{H} = \frac{\mathscr{J}}{2}\sum \boldsymbol{m}_i \cdot \boldsymbol{m}_j - \frac{\mathscr{K}}{2}\sum (\boldsymbol{m}_i \cdot \boldsymbol{\tau}_i)^2. \tag{S5}$$

Here $\mathscr{J} > 0$ is an effective exchange integral which favors AFM ordering, $\mathscr{K} > 0$ is an effective easy-tangential anisotropy constant, $j$ runs over nearest neighbors, and $\boldsymbol{\tau}_i = \boldsymbol{\tau}(ai)$ is a tangent unit vector to the wire. Note that the normal $\boldsymbol{\tau}_i$ introduces the information about the wire shape into the model (S5).

The length scale in simulations is defined with the magnetic length as $\ell = a\sqrt{\mathscr{J}/\mathscr{K}}$, the dimensionless time is defined as $\tau = 2\gamma_0 t\sqrt{\mathscr{J}\mathscr{K}}/\mu_s$.

The magnetization dynamics is simulated by means of numerical solution of the set of ODE (S4) for the initial conditions determined by the initial magnetization.

### 1. Simulations of parabola wire

We considered the parabola wire with geometry defined in (12). In simulations we considered parabola wire with $N = 1001$, and magnetic length $\ell = 20a$. The extreme curvature $\varkappa_0$ was varied in the range $\varkappa_0 \in [0, 0.2]$ with a step $\Delta\varkappa_0 = 0.01$.

The numerical experiment consist of two steps. Initially we relaxed the DW structure in an over-damped regime ($\eta = 0.5$) with initial position $q(0) = q_0 + 2\ell$ and phase $\Phi(0) = \Phi_0 + \pi/10$. In the second step we simulate a free dynamics of the system with low damping $\eta = 10^{-4}$. The equilibrium state of DWs are presented in Fig. 1(a) and 1(b). To determine the values of $q$ and $\Phi$ we extract the curvilinear components of Néel vector $n_{\text{T}} = \boldsymbol{n} \cdot \boldsymbol{e}_{\text{T}}$, $n_{\text{N}} = \boldsymbol{n} \cdot \boldsymbol{e}_{\text{N}}$, and $n_{\text{B}} = \boldsymbol{n} \cdot \boldsymbol{e}_{\text{B}}$ from simulation data, and apply fitting with Ansatz (4). The corresponding time evolution of DW position and phase is presented in Figs. 2(a) and 2(b), respectively, where one can see harmonic decaying oscillations with well pronounced frequencies $\omega_q$ and $\omega_\Phi$.

### 2. Simulations of Euler spiral wire

We considered the Euler spiral wire with geometry defined in (13). In simulations we considered wire with $N = 1001$, magnetic length $\ell = 20a$, and gradient of the curvature $\chi \in [1, 10] \times 10^{-3}$ with step $\Delta\chi = 1 \times 10^{-3}$.

Similarly to the case of parabola geometry, here we also performed simulations in two steps. Firstly, we relaxed DW structure in an overdamped regime ($\eta = 0.5$) with initial position $q(0) = 0$ and phase $\Phi(0) = \Phi_0$. The initial position for DW corresponds to the curvature with $\varkappa[q(0)] = 0$. In the second step we simulate a free dynamics of the system with low damping $\eta = 10^{-2}$. The analysis of the DW behavior was performed in the same manner as for parabola.

The velocity for DWs was obtained as fitting parameter of $q(\tau)$ obtained in simulations. For fitting we used trial functions $q(\tau) = q_0 + v\tau$.

### 3. Movies of the DW motion

For a better illustration of the DW motion induced by the curvature gradients, we prepared movies of DW dynamics. Movies are based on data obtained by means of the numerical simulations.

- `AFM_DW_parabola_k_0.1_eta_10e-4.mp4` shows the dynamics of the AFM DW in the parabolic wire with the $\varkappa_0 = 0.1$ and Gilbert damping $\eta = 10^{-4}$.

- `AFM_DW_euler_spiral_chi_5e-3_eta_10e-2.mp4` shows the dynamics of the AFM DW in the Euler spiral with the gradient of the curvature $\chi = 5 \times 10^{-3}$ and Gilbert damping $\eta = 10^{-2}$.


\end{document}